\documentclass[a4paper,fleqn,usenatbib]{mnras}

\usepackage{newtxtext,newtxmath}

\usepackage[T1]{fontenc}
\usepackage{ae,aecompl}

\usepackage{graphicx}	
\usepackage{amsmath}	
\usepackage{amssymb}	
\usepackage{soul}
\usepackage{nicefrac}
\usepackage{tabu}

\newcommand{\designation}{UAO}
\newcommand{\designations}{UAOs}
\newcommand{\kms}{$\mathrm{km}\,\mathrm{s}^{-1}$}
\DeclareMathOperator*{\argmax}{arg\,max}

\title[Kinematical age for 1I/'Oumuamua]{A kinematical age for the interstellar object 1I/'Oumuamua}

\author[F. Almeida-Fernandes \& H. J. Rocha-Pinto]{
F. Almeida-Fernandes,
H. J. Rocha-Pinto
\\
Universidade Federal do Rio de Janeiro -- UFRJ, Observat\'orio do Valongo, Ladeira Pedro Ant\^onio 43, 20080-090 Rio de Janeiro, RJ, Brazil\\
}

\date{Accepted 2018 August 10. Received 2018 August 8; in original form 2018 January 12}

\pubyear{2018}

\begin{document}
\label{firstpage}
\pagerange{\pageref{firstpage}--\pageref{lastpage}}
\maketitle

\begin{abstract}
1I/'Oumuamua is the first interstellar object observed passing through the Solar System. Understanding the nature of these objects will provide crucial information about the formation and evolution of planetary systems, and the chemodynamical evolution of the Galaxy as a whole. We obtained the galactic orbital parameters of this object, considering 8 different models for the Galaxy, and compared it to those of stars of different ages from the Geneva-Copenhagen Survey (GCS). Assuming that the galactic orbital evolution of this object is similar to that of stars, we applied a Bayesian analyses and used the distribution of stellar velocities, as a function of age, to obtain a probability density function for the age of 'Oumuamua. We considered two models for the age-velocity dispersion relation (AVR): the traditional power law, fitted using data from the GCS; and a model that implements a second power law for younger ages, which we fitted using a sample of 153 Open Clusters (OCs). We find that the slope of the AVR is smaller for OCs than it is for field stars. Using these AVRs, we constrained an age range of 0.01--1.87 Gyr for 'Oumuamua and characterized a most likely age ranging between 0.20--0.45 Gyr, depending on the model used for the AVR. We also estimated the intrinsic uncertainties of the method due to not knowing the exact value of the Solar motion and the particularities of 1I/'Oumuamua's ejection.

\end{abstract}




\section{Introduction}

The discovery of the interstellar object 1I/2017 U1 ('Oumuamua) by the Pan-STARRS survey \citep{Chambers+2016} in October, 2017 \citep{MPC2017a, MPC2017b}, and confirmation of its interstellar origin \citep{delaFuenteMarcos+delaFuenteMarcos2017}, sets a new field in astrophysics: the study of materials coming directly from other stellar systems. Crucial information about the Sun comes from the analysis of asteroids, like its chemical abundances \citep{Palme1988} and age with precision unmatched by any methods in the literature \citep{Bouvier+Wadhwa2010}. By analysing unbound asteroidal objects (\designation), like 'Oumuamua, we may be able to access this kind of precise information for other stellar systems as well.

\begin{figure*}
\centering
\includegraphics[scale=0.57]{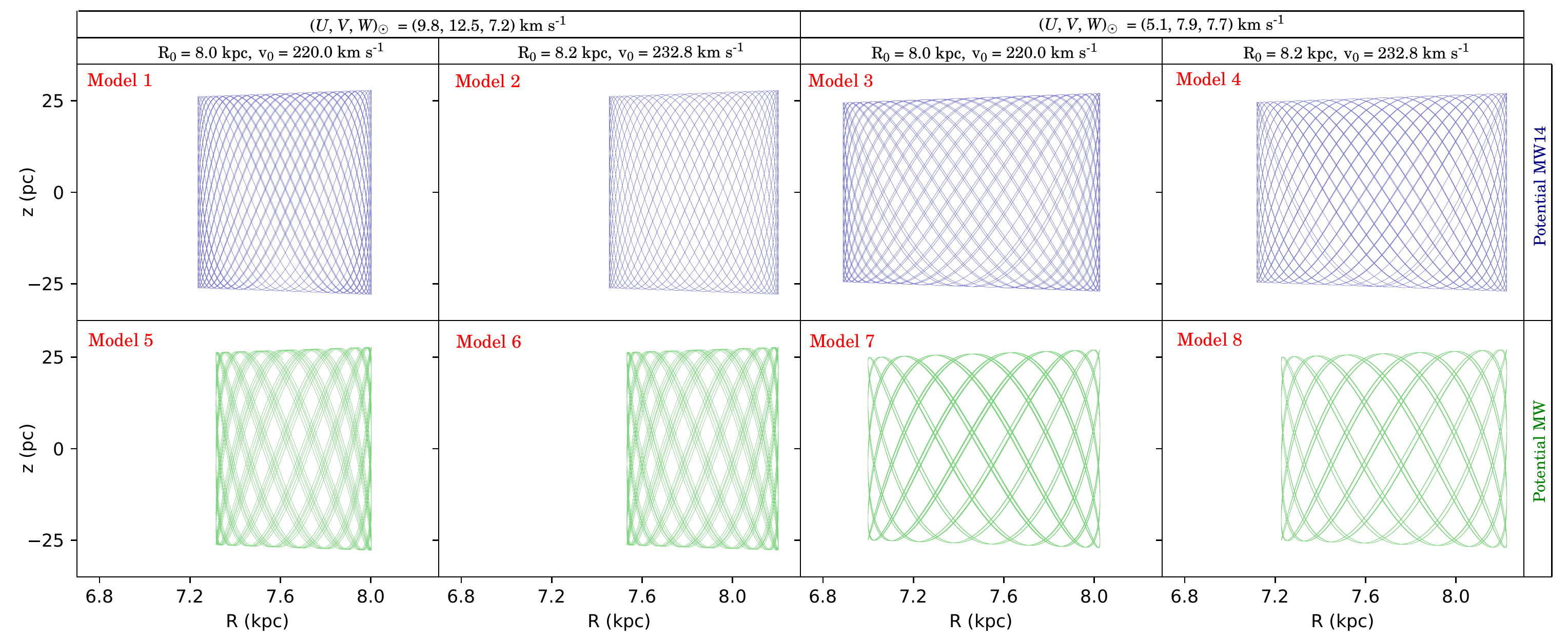}
\caption{'Oumuamua's Galactic orbit integrated in the past for several orbits ($\approx 3.5$ Gyr) for the eight models described in Table \ref{tab::orbital_parameters}. Although the shape of the orbit varies from one potential to the other, the orbital parameters ($e$, $z_\mathrm{max}$, $R_\mathrm{min}$ and $R_\mathrm{max}$) do not significantly change between them (in comparison to the range of these parameters observed for stars in the Solar Neighbourhood). Changing $R_0$ and $v_0$ does not affect the shape of the orbit, but causes a displacement. The model parameter that has the most significant effect in the derived orbital properties is the velocity of the Sun.}
\label{fig::orbits}
\end{figure*}

Current efforts are concentrated into determining how common these objects are, and if we can identify its original stellar system. \citet{PortegiesZwart+2017} take into account the volume sampled by the Pan-STARRS survey to estimate the number density of \designations \, and find the value of $7.0 \times 10^{14}$ pc$^{-3}$. They estimate a number of 2-12 encounters a year. Regarding the identification of its original system, \citet{Mamajek2017} finds that its spatial velocity is not compatible with any nearby star, while \citet{Gaidos+2017} suggest its origin to be the Carina and Columba associations (since its velocity differs by $\lesssim 2 \, \mathrm{km\,s}^{-1}$). \citet{Feng+Jones2017} go beyond the comparison of spatial velocities and integrate the orbits of 0.23 million local stars as well as that of 'Oumuamua and provide a list of 109 stars which had encounters closer than 5 pc (17 of which had encounters distances of less than 2 pc).

There is also interest in knowing the characteristics of the body, as it may provide further clues to its origin and provide constraints regarding the formation and evolution of planetary systems \citep[e.g.][]{Raymond+2017}. 'Oumuamua display an unusual elongated shape, with an axial ratio that may reach 10:1 and a rotational period of about 8 hr \citep{Knight+2017, Bolin+2017, Jewitt+2017}. Optical and spectral analysis show no evidence of cometary activity \citep{Ye+2017} but observed deviations in its predicted path indicates that some material is being expelled from the body \citep{Micheli+2018}. Its color is redder than asteroids but consistent with Kuiper Belt Objects \citep{Masiero2017}. \citet{Fitzsimmons+2017} suggested that its surface may be as organically rich as outer solar system bodies due to it displaying a similar red and featureless spectrum. However, one fundamental property of this body, that may provide clues about its nature and origin, remains unknown: the age. In this work, we aim to derive the age of this body from constraints provided by its galactic orbit.

The formation age of \designations\, is important for many reasons: (i) it helps narrowing down the list of potential stellar systems from which the object came from; (ii) combined with the chemical abundances observed in this and future similar objects, it will provide another tool to understand the chemical evolution of our galaxy; (iii) provide information about the dynamics and evolution of proto-planetary disks; (iv) and, considering that we can measure its $UVW$ velocity with higher precision than any star, these objects shed light to the whole dynamics of the Galaxy and the orbital heating mechanisms.

In this paper, we characterize the age of 'Oumuamua from a probability density function derived from the relation between stellar orbital parameters and age. Our methodology is explained in Section \ref{sec::methodology}. In Section \ref{sec::results} we show the results of the orbital integration and the obtained age. Finally, Section \ref{sec::conclusions} contains our conclusions.

\section{Methodology}
\label{sec::methodology}

It is known that the velocity dispersion of stellar groups increases with the stellar age \citep[e.g.][]{Wielen1977, Nordstrom+2004, Casagrande+2011, Martig+2014}. This is interpreted as a dynamical heating process of the stellar galactic orbits, caused by massive molecular clouds \citep[e.g.][]{Lacey1984, Hanninen+Flynn2002}, interactions with non-axissymetric Galactic structures like the bar \citep[e.g.][]{Saha+2010, Grand+2016} and the spiral arms \citep[e.g.][]{Carlberg+Sellwood1985, Martinez-Medina+2015} and/or mergers with satellite galaxies \citep[e.g.][]{Velazquez+White1999}.

\citet[][hereafter A\&R18]{AlmeidaFernandes+RochaPinto2018} take into account the relation between velocity distribution of stars, with respect to the Local Standard of Rest (LSR), and age. Through a Bayesian approach, A\&R18 used this relation to derive a probability density function (pdf) for the age of a star, based exclusively on its space velocity. Here, we discuss that this method, following the same relation as for stars, can be used for 'Oumuamua and other \designations.

For the kinematical method to work for \designations, the process of heating of the galactic orbit of these bodies must be the same as that for stars. It is expected that 'Oumuamua coalesced in the proto-planetary disk of a newly born star and initially shared the same galactic orbit as its parental star. Therefore, before the ejection of the body, this condition is straightforward: during this time, the changes in the \designation's galactic orbit will be exactly equal to that of the star. The continuation of the same heating process after the ejection needs a little more explanation: all the mechanisms considered to explain the disk heating (described above) involve collisions with bodies that are way more massive than the stars or the \designations. Therefore, as stellar masses are irrelevant during the collisions that cause the disk heating, the mass of \designations, being considerably lower, will be equally irrelevant, and the heating process is expected to be the same.

Even though the heating process is the same throughout the life of the \designation, we still have to consider the consequences of the ejection process in the change of its galactic orbit. If the ejection velocity is close to the escape velocity from the star, the excess velocity (the velocity between the \designation\, and its parental star at an infinite distance) will be close to zero, and the galactic orbit will not be significantly affected by the ejection. In this case the relation between the velocity distribution of \designations\, and their ages will be the same one observed for stars and the kinematical method can be applied. Otherwise, there is an additional ``kick'' that changes the velocity of the \designation. We approach this problem by applying the kinematical method as if the excess velocity was close to zero, and then, in Section \ref{sec::ejectionvelocity}, we investigate the effect of a non-zero excess velocity in the calculated ages.

\begin{figure}
\centering
\includegraphics[scale=0.59]{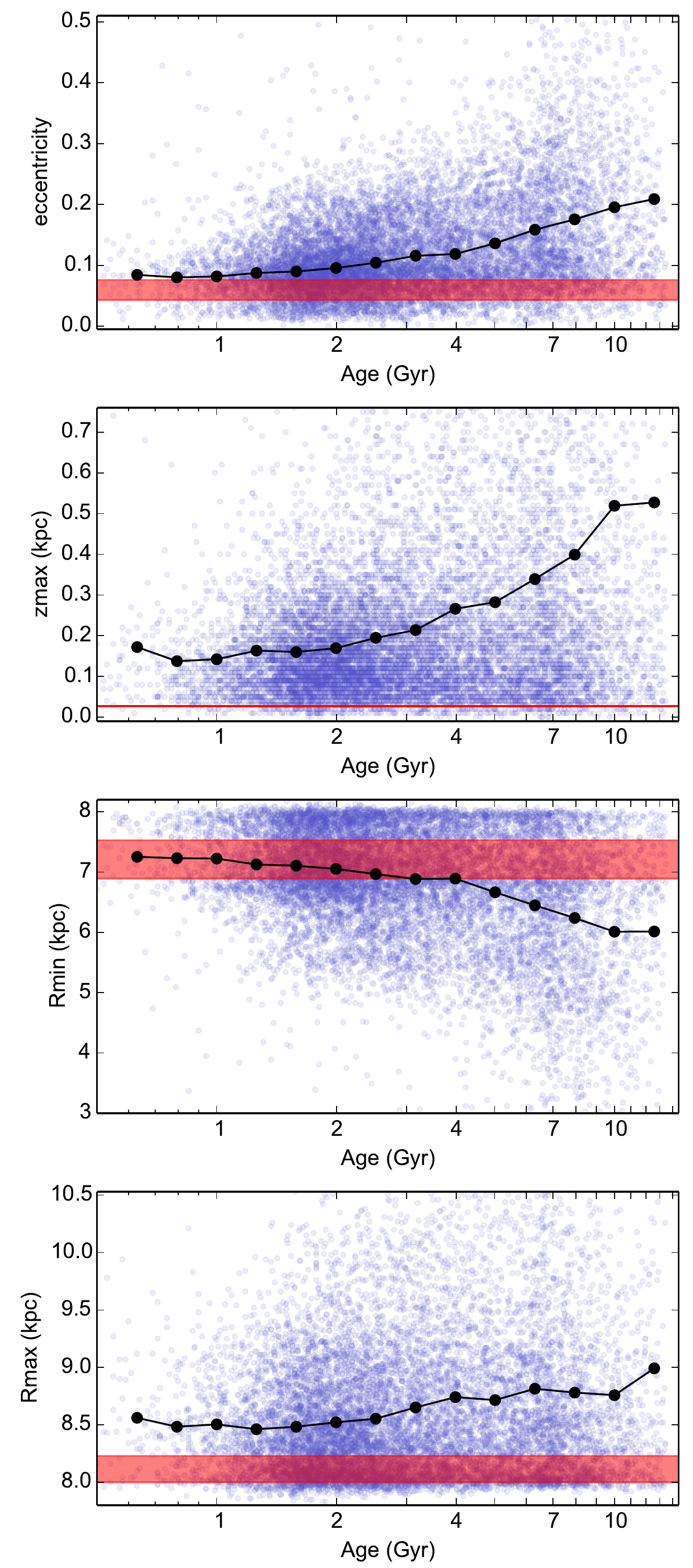}
\caption{Orbital parameters eccentricity ($e$), maximum distance to galactic plane ($z_\mathrm{max}$), perigalactic radius ($R_\mathrm{min}$) and apogalactic radius ($R_\mathrm{max}$) as a function of age for the stars of the GCS. For comparison, the values derived for 'Oumuamua are represented by the red shaded region, which extremes represent the minimum and maximum values obtained considering the eight galactic models described in Table \ref{tab::ages}}
\label{fig::parameters_age}
\end{figure}

Another working hypothesis that needs to be true in order to allow the application of the method is that the Sun is the first star visited by the \designation. The reason for this is that star-star interactions is a ruled out mechanism to explain the disk heating because stellar densities are too low. It has been estimated that collisions between stars can only significantly affect their orbits after a time of about $10^{14}$ Gyr \citep{Chandrasekhar1942}, which is greater than the age of the Universe, meaning that most stars will never go through such an encounter during their lifetimes. Considering that \designations\, have a lower collisional cross section, due to smaller masses, their probability of encountering a star is even lower, supporting our hypothesis that the Sun is the fist star visited by 'Oumuamua. In contrast, the high density of the \designations, as estimated by \citealp{PortegiesZwart+2018}, leads to an expected encounter rate between the Sun and \designations\, of up to a dozen per year. It is therefore expected that one \designation \, visits 0 to 1 star during its life time, while one star is visited by several \designations \, every year.

We can conclude that it is expected for the galactic orbit of \designations \, to evolve in the same way as that of stars due to the mechanisms of disk heating. Therefore, we can estimate the age of a \designation \, by comparing its kinematic properties with those observed for stars of different ages.

\section{Results and Discussion}
\label{sec::results}

\subsection{Orbital parameters and age}

We have backwards-integrated the orbit of 'Oumuamua using the orbital integrator \texttt{galpy} \citep{Bovy2015} and the velocities derived by \citet{Mamajek2017}: $(U,V,W) = (-11.457, -22.395, -7.746) \, \mathrm{km\,s}^{-1}$. To investigate how uncertainties in the Galactic model would affect the results, we have performed orbital integrations for two different galactic potentials: MWPotential and MWPotential2014, the default potentials of \texttt{galpy}. Both potentials are a combination of a \textit{power law with cut off} bulge, a \textit{Miyamoto-Nagai} disk and a \textit{Navarro-Frenk-White} halo. The main difference between them is the fact that MWPotential2014 was fitted using dynamical data of the Milky Way \citep[see][for more details]{Bovy2015}. We also used two sets of values for the Sun galactocentric distance and the LSR velocity for this radius and two different sets of values for the Sun peculiar velocities. The resulting orbits in the $zR$ plane are shown in Figure \ref{fig::orbits}. The parameters used in the integrations and the respectively calculated orbital parameters\footnote{In this work, the galactic orbital eccentricity is defined as $e = \nicefrac{(R_\mathrm{max}-R_\mathrm{min})}{(R_\mathrm{max}+R_\mathrm{min})}$.} are presented in Table \ref{tab::orbital_parameters}.

\begin{table*}
\centering
\caption{'Oumuamua's galactic eccentricity ($e$), maximum distance to galactic plane ($z_\mathrm{max}$), perigalactic distance ($R_\mathrm{min}$) and apogalactic distance ($R_\mathrm{max}$), obtained from orbital integrations using 8 different models of the Galaxy. We use: two standard Galactic potentials provided by \texttt{galpy}, MWPotential2014 and MWPotential; two sets of Solar velocities (A\&R18, \citealp{Koval+2009}); and two sets of Solar galactocentric distance ($R_0$) and LSR circular velocity ($v_0$). The extreme obtained values are displayed in boldface.}
\label{tab::orbital_parameters}
\begin{tabu} to \textwidth {llX[c]X[c]X[c]X[c]X[c]|X[c]X[c]X[c]X[c]}
\hline \hline
id & Potential & $U_\odot$ & $V_\odot$ & $W_\odot$ & $R_0$ & $v_0$ &    $e$ & $z_\mathrm{max}$ & $R_\mathrm{min}$ & $R_\mathrm{max}$ \\
   & & \kms & \kms & \kms & kpc & \kms & & pc & kpc & kpc \\ \hline
 1 & MWPotential2014 & 9.8 & 12.5 & 7.2 & 8.0 & 220.0 &  0.050 &  \textbf{•}textbf{27.91} &  7.23 &  8.00 \\
 2 & MWPotential2014 & 9.8 & 12.5 & 7.2 & 8.2 & 232.8 &  0.048 &  27.86 &  7.45 &  8.20 \\
 3 & MWPotential2014 & 5.1 & 7.9 & 7.7  & 8.0 & 220.0 &  \textbf{0.076} &  27.06 &  \textbf{6.89} &  8.03 \\
 4 & MWPotential2014 & 5.1 & 7.9 & 7.7  & 8.2 & 232.8 &  0.072 &  27.06 &  7.12 &  \textbf{8.23} \\
 5 & MWPotential & 9.8 & 12.5 & 7.2 & 8.0 & 220.0 &  0.045 &  27.75 &  7.31 &  \textbf{8.00} \\
 6 & MWPotential & 9.8 & 12.5 & 7.2 & 8.2 & 232.8 &  \textbf{0.043} &  27.71 &  \textbf{7.53} &  8.20 \\
 7 & MWPotential & 5.1 & 7.9 & 7.7  & 8.0 & 220.0 &  0.068 &  27.05 &  7.00 &  8.03 \\
 8 & MWPotential & 5.1 & 7.9 & 7.7  & 8.2 & 232.8 &  0.065 &  \textbf{27.05} &  7.23 &  8.23 \\ \hline \hline
\end{tabu}
\end{table*}

In Figure \ref{fig::parameters_age} we compare the orbital parameters derived for 'Oumuamua (the red shaded region contains the values derived considering all the eight models) with those of 12337 stars from the Geneva-Copenhagen Survey \citep{Nordstrom+2004, Casagrande+2011} with different ages (The black line represents the mean values as a function of age). 

For stars, we can see a clear relation between the distribution of the parameters and the age: older stars are more likely to display a non-circular and non-planar orbit, which reflects the dynamical heating of the disk. Therefore, higher values of galactic eccentricity, $z_\mathrm{max}$ and $R_\mathrm{max}$ and lower values of $R_\mathrm{min}$ can be used as a direct indication of older ages for \designations. In the case of lower values (higher for $R_\mathrm{min}$) for the kinematic parameters this interpretation is not so straightforward, as stars are observed with these parameters for all ages. 'Oumuamua's kinematical parameters lie in this region that is shared by stars of all ages, even considering 8 different models during the orbital integration. Therefore, we cannot infer any information about the age of 'Oumuamua from this preliminary analysis and a more complex statistical approach has to be applied.

\begin{figure*} 
\centering
\includegraphics[scale=0.57]{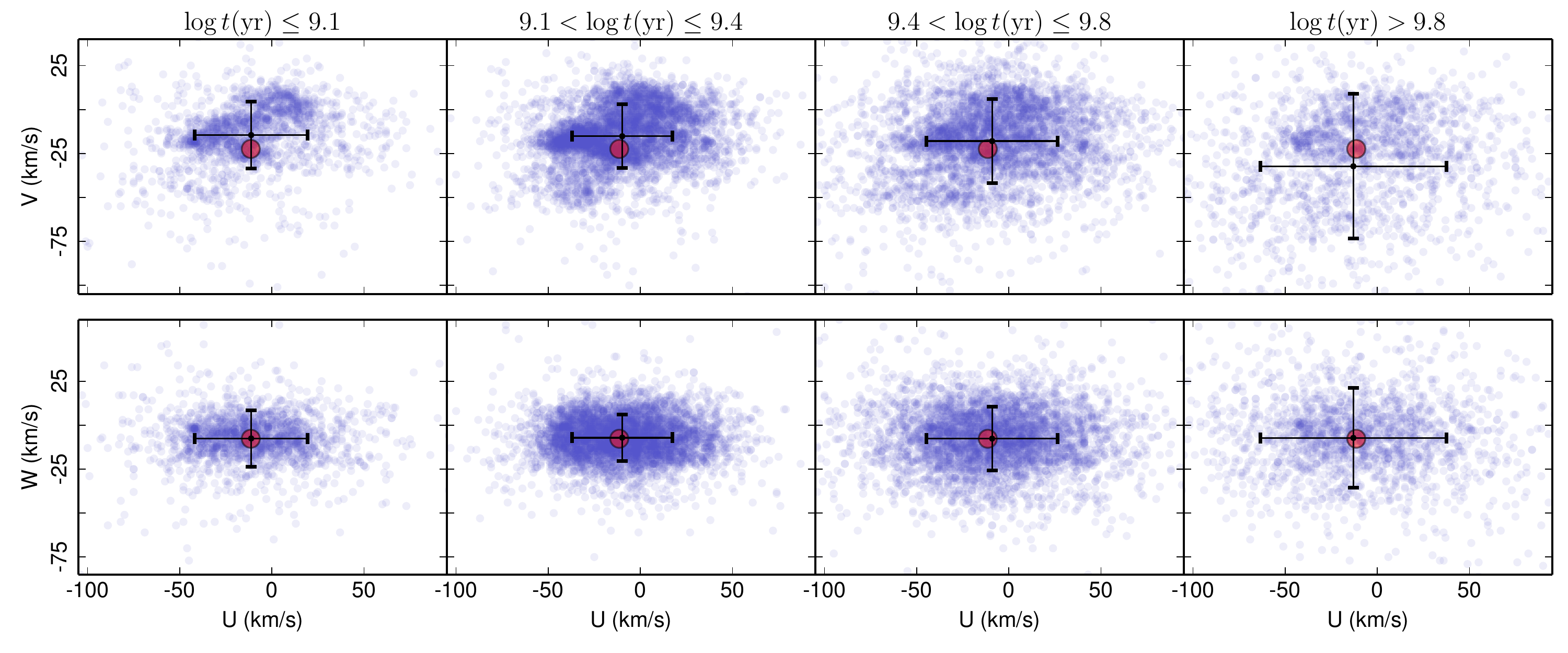}
\caption{Velocity distribution in the UV (left) and UW (right) planes for the stars of the GCS separated by age: $\log t(\mathrm{yr}) \leq 9.1$ (top), $9.4 < \log t(\mathrm{yr}) \leq 9.7$ (middle) and $\log t(\mathrm{yr}) > 9.8$ (bottom). The black crosses represent the mean and dispersion for each distribution. The velocity of 'Oumuamua is represent as the red circle for comparison. The increase of velocities dispersion as a function of age is evident, as well as the decrease of the mean $V$ velocity.}
\label{fig::UV_UW_dens}
\end{figure*}

\subsection{Kinematical method}

We have seen in Figure \ref{fig::parameters_age} that stars of all ages can have orbits close to a circular and planar one, but, as we consider older stars, it becomes more and more common to find stars on more eccentric and non-planar orbits. It means that the distribution of these orbital parameters is changing with age, which causes the probability for the star to display an orbital parameter to also change with age. The same effect can be seen considering the velocity space. In Figure \ref{fig::UV_UW_dens} we plot the heliocentric $UV$ and $UW$ velocity distribution\footnote{In this work, we consider that the positive direction of the U axis points towards the Galactic Center.} for the GCS stars of four different age groups: $\log t(\mathrm{yr}) \leq 9.1$ (left); $9.1 < \log t(\mathrm{yr}) \leq 9.4$ (middle-left); $9.4 < \log t(\mathrm{yr}) \leq 9.8$ (middle-right); and $\log t(\mathrm{yr}) > 9.8$ (right). The black dot and lines represents respectively the mean and standard deviation of the distributions. The velocity of 'Oumuamua is plotted for comparison as a red circle.

A\&R18 parametrize this distribution as a function of age using data from the GCS. The parametrization takes into account the increase of the velocity dispersion as a function of age, the correlation between the $U$ and $V$ velocities, possibly caused by the existence of moving groups, and the decrease of the mean $V$ as a function of age, which is an effect of the asymmetric drift. This parametrization as a function of age allows one to calculate the probability of a star (or in this case a \designation) to have a specific observed velocity: $p(U, V, W|t)$. Through a Bayesian approach, we can use this to derive a probability density function for the star:
\begin{equation}
p(\log t| U, V, W) \propto p(U, V, W| \log t) \cdot p(\log t) \,,
\end{equation}
where $p(\log t)$ is the prior distribution function and concentrate the information we have about the age before considering the kinematic parameters. In this case, we only consider that the maximum age cannot exceed 13.8 Gyr (the age of the Universe), and use a flat prior (in the variable $t$, instead of $\log t$) for lower ages. This means that we are not considering any other information to derive the age besides the spatial velocity and the age of the Universe, but can also be physically interpreted as assuming a constant star formation rate for the Galaxy. In this paper, we have chosen to work with $\log t$ instead of $t$ because we observed that the derived pdf changes much more slowly for older ages than for younger ages. For this reason, uniform changes in $t$ lead to very different changes in probability for younger and older ages, while uniform changes in $\log t$ lead to a more symmetrical distribution that favours the characterization of the age and estimation of uncertainties.

The kinematical method is explained with more details in A\&R18, and here we give just a brief summary of it. The probability $p(U, V, W| \log t)$ is calculated applying the formalism of the velocity ellipsoid \citep{Binney+Merrifield1998}. The velocity distribution is modelled as a multivariate normal distribution with zero mean and dispersion depending on age. To describe the velocity distribution by this model, we first need to apply a transformation to the velocities taking into account the non-zero Solar motion and the time-dependent correlation between the $U$ and $V$ components:
\begin{subequations}
\label{eq:v1v2v3}
\begin{align}
v_1 &= (U+U_{\sun}) \, \cos{\ell_v} + (V+V'_{\sun}) \, \sin{\ell_v} \,, \\
v_2 &= -(U+U_{\sun}) \, \sin{\ell_v} + (V+V'_{\sun}) \, \cos{\ell_v} \,, \\
v_3 &= W+W_{\sun} \,,
\end{align}
\end{subequations}
and
\begin{equation}
\ell_v = \frac{1}{2}\arctan{\left( \frac{2\,\sigma_{UV}}{\sigma^2_U-\sigma^2_V} \right)}\,.
\end{equation}
where $V'_{\sun}$ is a combination of the constant $V$ component of the Solar motion ($V_{\sun}$) and the time dependent asymmetric drift, and is parametrized as $V'_{\sun} = V_{\sun} + a\,t + b\,t^2$. The vertex deviation, $\ell_v$, is also time dependent. Finally the probabilities are given by:
\begin{equation}
p(U, V, W|\log t) = p(v_1|\log t) \cdot p(v_2|\log t) \cdot p(v_3|\log t)
\end{equation} 
The probability for each component, $v_i = v_1, v_2, v_3$, is then given by:
\begin{equation}
p(v_i|\log t) = \frac{1}{(2\,\pi)^{\nicefrac{1}{2}} \, \sigma_i(\log t)} \exp{ \left(-  \frac{v_i^2}{2 \sigma_i(\log t)^2}\right)}
\end{equation}

A\&R18 parametrized the time dependent parameters ($V'_{\sun}$, $\ell_v$ and $\sigma_i$) using data from the GCS, allowing the calculation of $p(\log t| U, V, W)$ for different ages and, therefore, the construction of an age probability density function (pdf).

\begin{figure}
\centering
\includegraphics[scale=0.57]{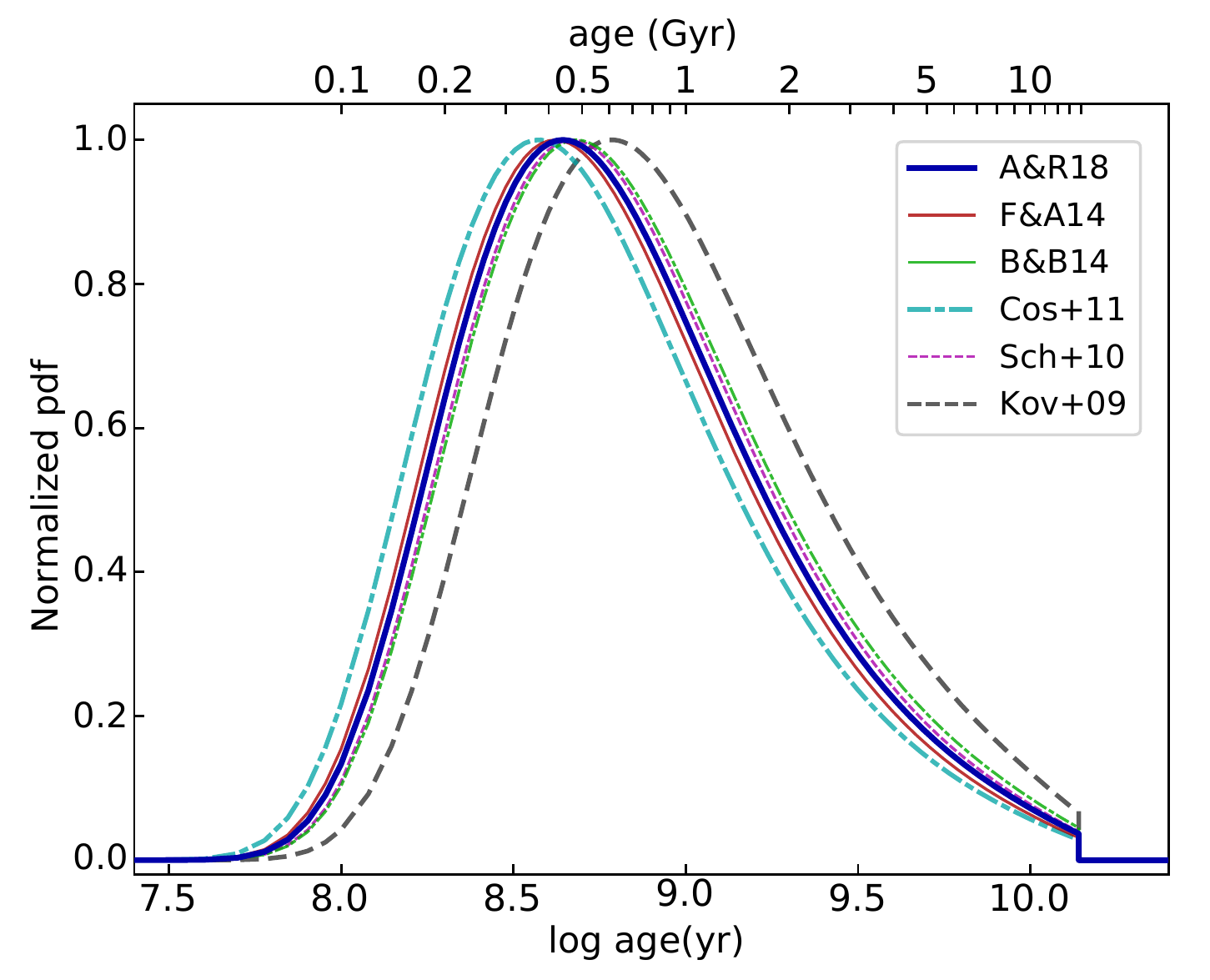}
\caption{Age pdfs obtained through the kinematical method for the \designation \, 'Oumuamua considering six different sets of Solar velocities (described in table \ref{tab::ages}). Except for $UVW_\odot$(Kov+09) and $UVW_\odot$(Co{\c s+11}), there is negligible difference between the obtained pdfs.}
\label{fig::pdfs}
\end{figure}

Using the velocity of 'Oumuamua, prior to its encounter with the Sun, we derived an age pdf using the method described above. To take into account the uncertainty in the Solar motion, we have calculated the pdf for six different sets of $UVW_\odot$ found in the literature. The resulting pdfs are shown in Figure \ref{fig::pdfs}. From the pdf, the age can be characterized from the most likely age ($\log t_\mathrm{ML}$) and expected age ($\log t_\mathrm{E}$):
\begin{subequations}
\begin{align}
\log t_{\mathrm{ML}} &= \argmax_t f(\log t) \\
\log t_{\mathrm{E}} &= \int\limits_{-\infty}^{\infty} \log t \, f(\log t) \, \mathrm{d}\log t
\end{align}
\end{subequations}
where $f(\log t)$ represents $p(\log t|U, V, W)$.

We also calculate the spread of the pdfs using the $2.5\%$, $16\%$, $84\%$ and $97.5\%$ percentiles, which can be used as an indication of the uncertainty of the age determination. The result for all values of Solar peculiar velocities are shown in Table \ref{tab::ages}. Removing the two extreme values and taking the average of the results for the remaining Solar motions, we obtain $\log t_\mathrm{ML} = 8.65$ ($t_\mathrm{ML} = 0.45$ Gyr) and $\log t_\mathrm{E} = 8.82$ ($t_\mathrm{E} = 0.66$ Gyr). Both age estimators are affected by different bias: the most likely age is biased towards the extremes of the age interval, while the expected age is biased towards the center of the interval. Comparing data of the GCS with Isochronal determined ages A\&R18 suggest a weighted average between $t_\mathrm{ML}$ and $t_\mathrm{E}$ that allows the minimization of the bias from both ages. For 'Oumuamua, this age results in $t_\mathrm{kin} = 0.50$ Gyr. Using the 16\% and 84\% percentiles as a proxy for $\sigma_t$, we finally obtain $t_\mathrm{kin} = 0.50^{+1.37}_{-0.27}$ Gyr.

\begin{table*}
\centering
\caption{Ages characterized from the pdfs obtained through the kinematical method. We have considered six sets of Solar velocities and derived the most likely age, expected age and age percentiles for each case. The extreme values are highlighted in boldface and were excluded from the analyses.}
\label{tab::ages}
\begin{tabu} to \textwidth {X[c]X[c]X[c]X[6]|X[c]X[c]X[c]X[c]X[c]X[c]X[c]}
\hline \hline
 & & & & & & \multicolumn{5}{c}{$\log t$ percentiles} \\
$U_\odot$ & $V_\odot$ & $W_\odot$ & Reference & $\log t_\mathrm{ML}$ & $\log t_\mathrm{E}$ & 2.5\% & 16\% & 50\% & 84\% & 97.5\% \\ \hline 
 9.8 & 12.5 & 7.2 & A\&R18                                & 8.64 & 8.81 & 8.04 & 8.35 & 8.75 & 9.26 & 9.80 \\
14.1 & 14.6 & 6.9 & \citet{Francis+Anderson2014} (F\&A14) & 8.62 & 8.78 & 8.03 & 8.33 & 8.72 & 9.23 & 9.78 \\
 6.0 & 10.6 & 6.5 & \citet{Bobylev+Bajkova2014} (B\&B14) & 8.68 & 8.85 & 8.08 & 8.38 & 8.79 & 9.31 & 9.83 \\
 8.8 & 14.2 & 6.6 & \citet{Coskunoglu+2011} (Co{\c s}+11) & \textbf{8.58} & \textbf{8.74} & \textbf{7.98} & \textbf{8.28} & \textbf{8.68} & \textbf{9.19} & \textbf{9.75} \\
11.1 & 12.2 & 7.2 & \citet{Schonrich+2010} (Sch+10) & 8.66 & 8.83 & 8.07 & 8.37 & 8.77 & 9.29 & 9.82 \\
 5.1 &  7.9 & 7.7 & \citet{Koval+2009} (Kov+09) & \textbf{8.78} & \textbf{8.94} & \textbf{8.17} & \textbf{8.48} & \textbf{8.89} & \textbf{9.41} & \textbf{9.89} \\ \hline \hline
\end{tabu}
\end{table*}

The ages obtained using different solar motions also allows us to check how this uncertainty translates to the derived age uncertainty. The standard deviation of the derived values is 0.07 Gyr for the most likely age and 0.10 Gyr for the expected age. Therefore, considering the weighted average, the derived kinematical age has an additional uncertainty of about 0.08 Gyr caused by our lack of knowledge of the exact value for the Solar motion.

\subsection{Uncertainty caused by the ejection velocity}
\label{sec::ejectionvelocity}
One of the assumptions we made to derive the kinematical age was that the galactic orbit of 'Oumuamua would not be significantly different from that of its parental star. Here we test how robust the method is by considering different escape velocities and checking how much they affect the derived age.

The velocity that we have to consider for the kinematical method is the hyperbolic excess velocity, $v_\infty$, which is the velocity of the body, in relation to the parental star, as the distance between them tends to infinity. The excess velocity is related to the escape velocity $v_{e}$ and the velocity, $v$, of the body at a given point of its orbit by the relation:
\begin{equation}
v_{\infty}^2 = v^2 - v_{e}^2
\label{eq::velocity_excess}
\end{equation}

It follows from Equation \ref{eq::velocity_excess} that the excess velocity will be zero in the case the body escapes with velocity $v = v_e$ and therefore will follow a very similar galactic orbit as its parental star. This is the ideal case, in which a body like 'Oumuamua will experience the same subsequent disk heating as the field stars. In the cases the body escapes with a velocity higher than the escape velocity, the body suffers a initial ``kick'' before the disk heating starts to act on its galactic orbit. This initial difference between the spatial velocity of the parental star and the ejected body will affect its age determination.

We check how much a non-zero excess velocity affects the age determination considering two different cases: (i) $v_{\infty} = 5$ \kms and (ii) $v_{\infty} = 10$ \kms. These values were chosen after calculating the excess velocities for different scenarios considering different semi-major axes for the body's orbit around its parental star and different stellar masses. For instance, in a Sun-like star, for a body in an asteroid belt-like population (semi-major axis of around 2 AU), an excess speed of $v_{\infty} = 10$ \kms$\,$ is only achieved if the body reaches a kinetic energy that is 10--20\% higher than the one necessary to be ejected from the system. When considering objects in a Kuiper belt-like population (semi-major axis of around 40 AU), around a Sun-like star, an excess velocity of $v_{\infty} = 5$ \kms$\,$ can only be achieved if the kinetic energy at the time of ejection is about two times larger than the kinetic energy of the escape velocity. Therefore, we consider these two ejection excess velocities as typical values of what can be expected for bodies like 'Oumuamua.

In order to investigate how the excess velocity affects the age determination we generate a sample of 10000 ($U$, $V$, $W$) velocities sets, whose total velocity differs from the 'Oumuamua's measured velocity by an amount equal to the chosen velocity excess. The directions are isotropically sampled and the velocity set represents the possible velocities 'Oumuamua's parental star may have considering this velocity excess. The kinematical method is applied and the ages are derived for each of the velocities sets.

\begin{figure}
\centering
\includegraphics[scale=0.58]{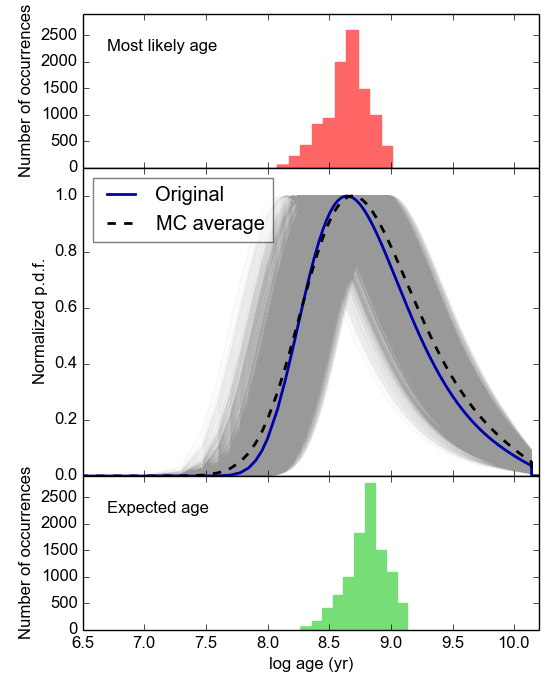}
\caption{Age pdfs calculated for the 10000 re-samples of the $UVW$ velocities considering an excess velocity of 5 \kms$\,$ for the ejection of 'Oumuamua (middle panel). The dashed line represents the average of all the Monte Carlo pdfs. The original age pdf is also shown in blue for comparison. The top and bottom panel represent the distribution of the characterized most likely and expected ages, respectively.}
\label{fig::deltav5dist}
\end{figure}

Figure \ref{fig::deltav5dist} shows the calculated age pdfs for the case of $v_{\infty} = 5$ \kms and the average between all curves (dashed black line). For comparison, the original derived pdf is represented in solid blue line. We also show the distribution of the obtained most likely ages (top panel) and expected ages (lower panel). The average values for the most likely and expected ages are 0.47 and 0.67 Gyr, with standard deviations of 0.17 and 0.23 Gyr, respectively. This values are slightly higher than those derived considering 'Oumuamua's measured space velocity (which is expected considering the derived result lies towards the lower side of the age limit), but are still within the uncertainties. The spread of the point estimators distribution allows us to characterize the possible bias in the age estimators due to the unknown parental star velocity for the case of an excess velocity of 5 \kms, which we found to be of about 0.2 Gyr. By comparing the spread of the averaged curve to that of the original pdf, we can estimate the contribution caused by not knowing the ejection velocity to the derived age uncertainty. In this case we find that the spread, defined as the difference between the 16 and 84 percentiles, increases by 8\%.

\begin{figure}
\centering
\includegraphics[scale=0.58]{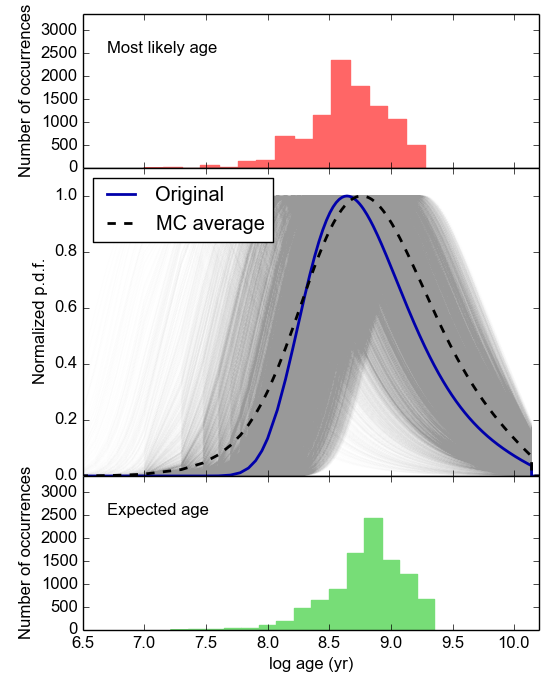}
\caption{The same as Figure \ref{fig::deltav5dist}, but for an excess velocity of 10 \kms for the ejection of 'Oumuamua.}
\label{fig::deltav10dist}
\end{figure}

We repeated the analysis for the case of excess velocity of 10 \kms$\,$ and show the results in Figure \ref{fig::deltav10dist}. In this case, we find that the average most likely and expected ages are 0.55 and 0.76 Gyr, with standard deviations of 0.37 and 0.46, respectively, still slightly biased towards higher values than previously measured but also within the predicted uncertainties. For this scenario, the estimated uncertainty increases by 22\%.

From this analysis, we can conclude that the point estimators may be slightly biased up to $\approx$ 0.2--0.4 Gyr and an additional uncertainty of $\approx$ 10--20\%, imposed by unknown characteristics of 'Oumuamua's ejection, must be considered for the estimated age. 

\subsection{Extended analysis for younger ages}

Stars originate from molecular clouds together with other thousands stars, which remain grouped in their early stages of life. This group of stars form what is known as open clusters (OCs) or stellar associations. These stars are gravitationally bound and follow the same orbit in the Galaxy. In other words, it means their $U$, $V$, $W$ velocities are very similar, which contradicts the random velocity distribution assumed by the kinematical method. 
'Oumuamua's host star was likely part of an OC in its earlier stages of life and we can take this into account to improve our age determination.

The kinematical method described in \citet{AlmeidaFernandes+RochaPinto2018} parametrizes the velocity distribution using data from the Geneva-Copenhagen Survey \citep{Casagrande+2011}. More than 90\% of the stars in the survey are estimated to be older than 1 Gyr. Therefore, the probabilities obtained for younger ages are calculated from an extrapolated velocity distribution. One way to remedy this is to consider that younger stars are still part of an OC or stellar association.

As for stellar groups, the spatial velocity dispersion of a group of OCs is known to increase with its mean age \citep{Wu+2009}. However, the exact relation between velocity dispersion and age may be different between OCs and field stars. \citet{VandePutte+2010} calculated the $U$, $V$, $W$ velocities of 488 OCs from the DAML catalogue \citep{Dias+2002} to investigate their galactic orbits. Here we use the velocities and the age for the OCs, also provided in the catalogue, to fit the relation between velocity dispersion and age for OCs and discuss how to include the result in the methodology of the Kinematical Method.

Before fitting the age--velocity dispersion relation (AVR), we removed from the sample the OCs with uncertainties higher than 5 \kms$\,$ for the $U$ and $W$ components and higher than 20 \kms$\,$ for the $V$ component. This step is necessary to ensure that the measured velocity dispersion is real and not an effect of observational errors. The reason for the relaxed criteria for the $V$ component is that, in the catalogue, this velocity is coupled with the velocity of the LSR and also incorporates its uncertainty. Of the remaining 178 OCs, 158 had age data provided in the catalogue. Finally, as our goal is to focus on younger OCs, we removed 5 OCs whose ages were higher than 1 Gyr.

We have divided the final sample of 153 OCs in three groups according to their ages and calculated the mean age and velocities dispersions for each group. In order to apply the kinematical method, we also have to calculate the velocity dispersion for the principal components of the velocity ellipsoid \citep{Binney+Merrifield1998}, designated as $\sigma_1$ and $\sigma_2$. The values of $\sigma_1$ and $\sigma_2$ depend on the value of the vertex deviation $\ell_v$, which is a measurement of the correlation between the $U$ and $V$ components. For stars, A\&R18 show that the vertex deviation is non-zero and decreases with the mean age of a stellar group. In the case of OCs, we find that the correlation between the $U$ and $V$ velocities is very small and can be neglected. In this case, we have $\ell_v = 0$ and the dispersions $\sigma_1$ and $\sigma_2$ are equal to $\sigma_U$ and $\sigma_V$ respectively.

\begin{figure}
\centering
\includegraphics[scale=0.9]{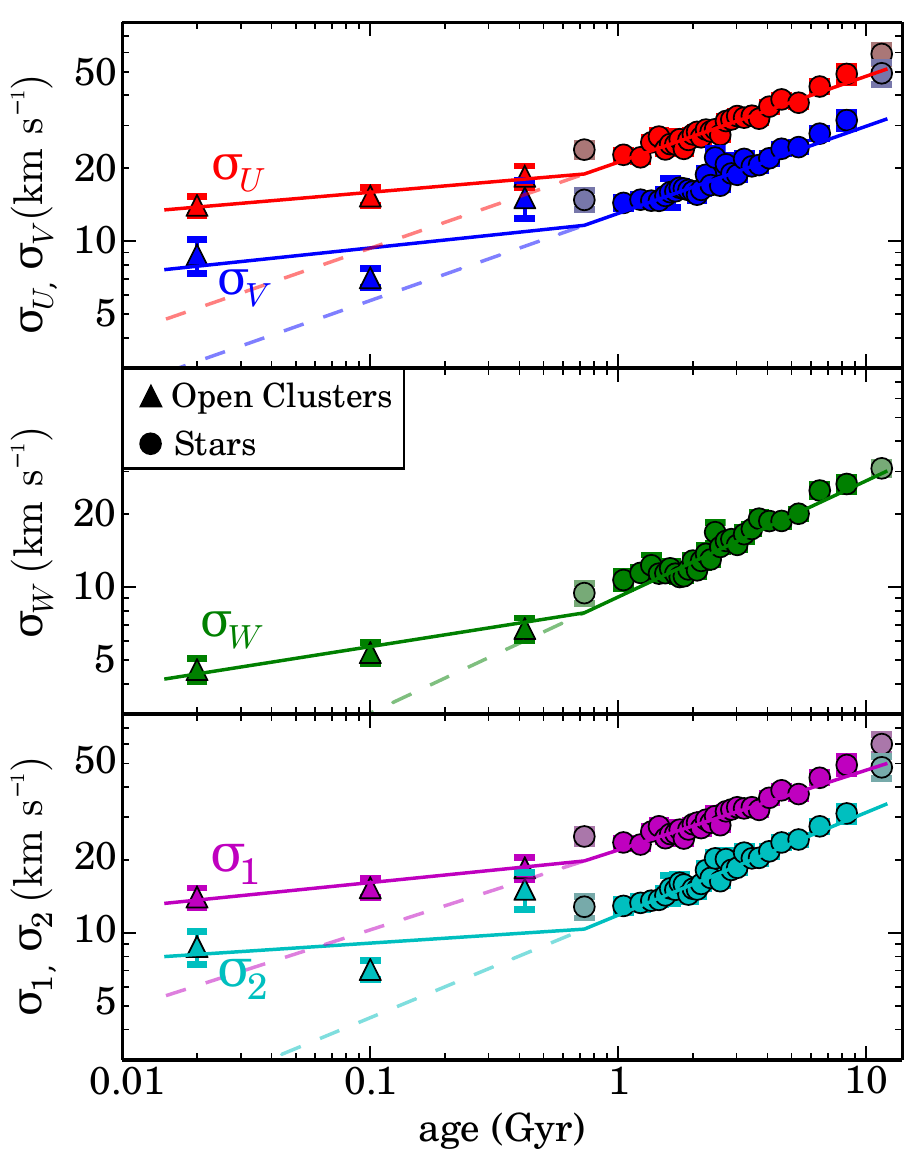}
\caption{Velocity dispersion as a function of age for Open Clusters (triangles) and field stars (circles) for the $U$ (top panel, red), $V$ (top panel, blue), $W$ (middle panel, green) components and for the principal components of the velocity ellipsoid $v_1$ (bottom panel, magenta) and $v_2$ (bottom panel, cyan). The solid lines represents the fitted relations parametrized as in Equation \ref{eq::sigma}. The dashed line represents the extrapolation for younger ages of the relation obtained for field stars.}
\label{fig::sigmafit}
\end{figure}

Figure \ref{fig::sigmafit} show the results obtained for the OCs as well as the points calculated by A\&R18 for field stars using data from the Geneva-Copenhagen Survey. We find that, similar to field stars, the velocity dispersions increase with the mean age of the OCs groups, confirming the result of \citet{Wu+2009}. However, the increase is considerably stepper in the case of field stars. Therefore, in order to fit a relation that satisfies both the OCs and the field stars data, we chose to fit a different power law for each regime. We also impose the condition that the relation must be continuous, and arbitrarily chose the age of 0.73 Gyr as the separation between both regimes. This age was chosen with the goal of minimizing the need of extrapolation for the fitted model, since this is the younger age for which we have data for field stars. The relation between velocity dispersion and age can then be written as:
\begin{equation}
\sigma_i(t) = \begin{cases}
	b_i\,t^{a_i} = k_i\left(\frac{t}{t_0}\right)^{a_i}, & \mathrm{if} \; t \geq t_0 \\
	k_i\left(\frac{t}{t_0}\right)^{c_i}, & t < t_0
\end{cases}
\label{eq::sigma}
\end{equation}
where $i$ represent each component $U$, $V$, $W$, $v_1$ and $v_2$, $k_i = b_i \,t_0^{a_i}$ and we have set $t_0$ as 0.73 Gyr.

The obtained parameters $a_i$, $b_i$, $c_i$ and $k_i$ are shown in Table \ref{tab:fit_sigma}. We note that the $a_i$ and $b_i$ parameters are exactly the same as calculated in A\&R18, since it corresponds to the same data for field stars. The fitted $\sigma(t)$ relations are plotted in Figure \ref{fig::sigmafit} as solid lines. The dashed line represents the extrapolation for younger ages of the relation fitted for field stars and show that it considerably under-predicts the dispersion for OCs.

\begin{table}
	\centering
	\caption{Parameters obtained for the model of the AVR for each velocity component, including $v_1$ and $v_2$. The model used to fit these parameters is described by Equation \ref{eq::sigma}.}
	\label{tab:fit_sigma}
	\begin{tabular}{ccccc} 
		\hline
		 & $a_i$ & $b_i$ & $c_i$ & $k_i$\\
		\hline
		$\sigma_U$ & $0.354$ & $21.18$ & $0.087$ & $18.95$\\
		$\sigma_V$ & $0.359$ & $13.03$ & $0.108$ & $11.64$\\
		$\sigma_W$ & $0.479$ & $9.11$ & $0.161$ & $7.84$\\
		$\sigma_1$ & $0.329$ & $21.98$ & $0.103$ & $19.82$\\
		$\sigma_2$ & $0.423$ & $11.85$ & $0.067$ & $10.37$\\
		\hline
	\end{tabular}
\end{table}

The difference between the observed slopes of the AVR for OCs and field stars indicates that OCs are more resistant to the disk heating. This could be explained by their difference in total mass and degrees of freedom. OCs masses can be 3--4 orders of magnitude greater than that of single stars, which can be non-negligible when compared to the masses of the agents causing the disk heating. The OCs also have more degrees of freedom to distribute the energy received through gravitational encounters, which, instead of causing an increase of the global spatial velocity of the cluster, may be used to change the internal stellar distribution or even cause the ejection of stars. Therefore, it can indeed be expected that OCs will be more resistant to the mechanisms responsible to the disk heating, causing the slope of the AVR relation to be smaller. We claim, however, that a thorough investigation of the differences between the AVR of OCs and field stars is beyond the scope of this work and here we simply use the fitted AVR to model the velocity distribution and apply the kinematical method.

\begin{figure}
\centering
\includegraphics[scale=0.45]{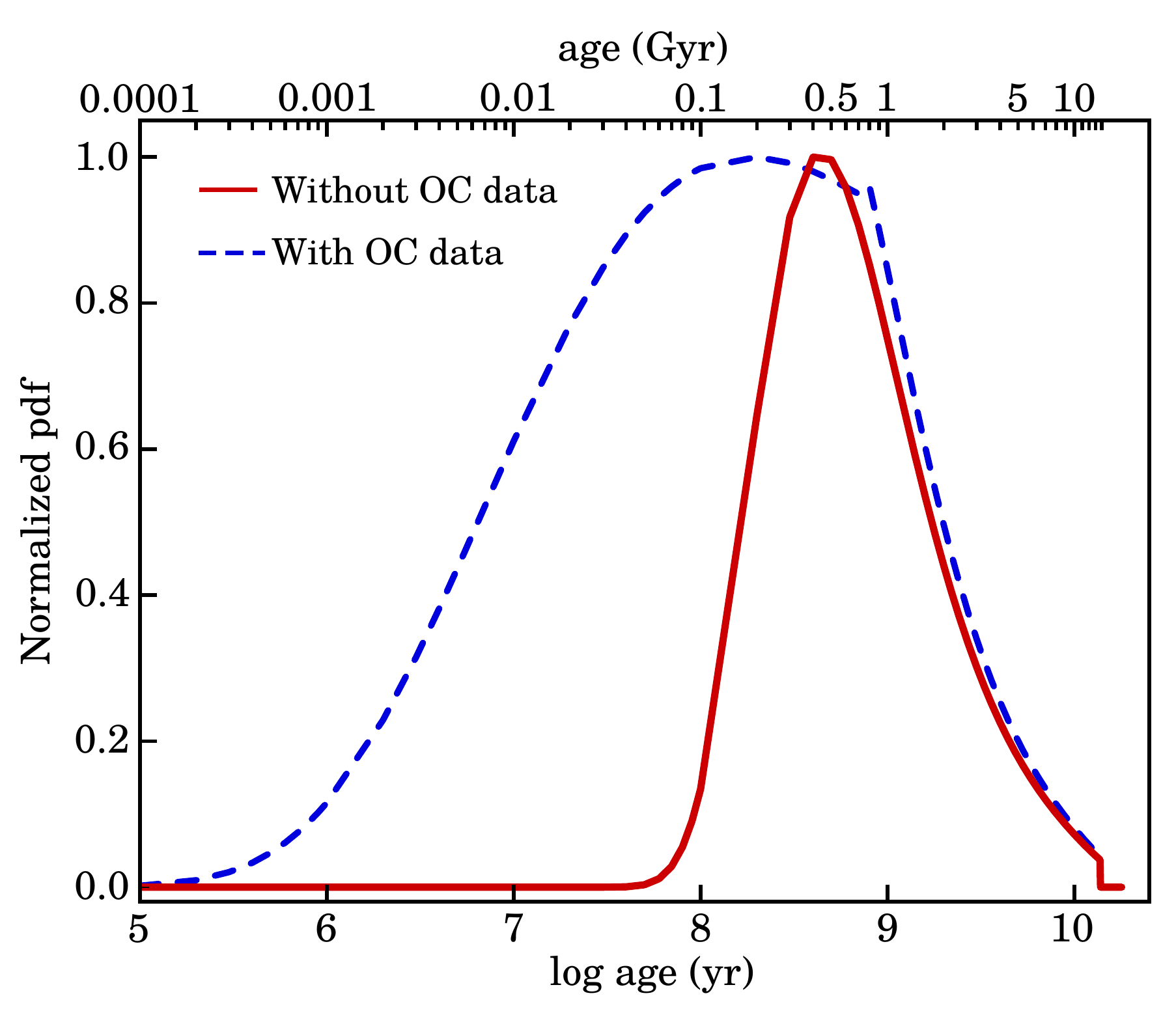}
\caption{Age pdf obtained using the AVR of OCs for younger ages (dashed, blue line), and using the extrapolated AVR of field stars for younger ages (solid, red line).}
\label{fig::age_pdf_with_OC}
\end{figure}

We then recalculate the age pdf considering that for younger ages, the host star of 'Oumuamua was part of a OC and was latter ejected following the traditional AVR for field stars. Figure \ref{fig::age_pdf_with_OC} shows a comparison between the age pdfs obtained considering the OCs data (blue dashed line), and the one obtained simply by extrapolating the AVR of field stars (red solid line). For higher ages, the age pdf is basically unchanged, but considerable differences can be observed for lower ages. As the AVR is shallower for the OC scenario, the change in the velocities distribution for different ages is smaller, causing the changes in the calculated probabilities to be less significant. Therefore, the lower limit for 'Oumuamua's age is less constrained in this case. We obtained the values of $\log t_\mathrm{ML} = 8.30$ ($t_\mathrm{ML} = 0.20$ Gyr) and $\log t_\mathrm{E} = 8.05$ ($t_\mathrm{E} = 0.11$ Gyr). Again, considering the weighted average proposed by A\&R18, we obtain $t_\mathrm{kin} = 0.18^{+0.70}_{-0.17}$ Gyr.

\begin{figure}
\centering
\includegraphics[scale=0.53]{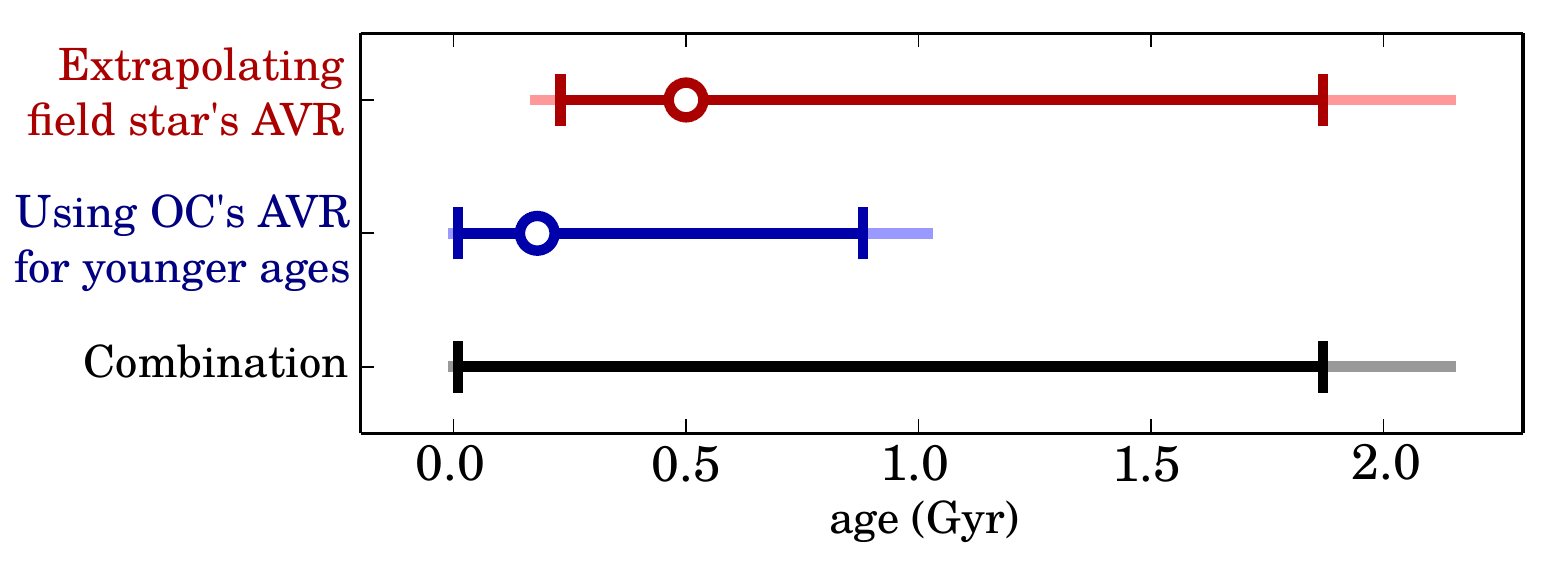}
\caption{Characterized kinematical ages as well as their lower and upper limits for the scenarios using the extrapolated AVR of field stars (top, red), using the AVR of OCs for younger stars (middle, blue) and the combination of both results (bottom, black). The faded lines include an addition 20\% uncertainty estimated from not knowing the exact conditions of 'Oumuamua ejection.}
\label{fig::age_comparison}
\end{figure}

Since we cannot know when 'Oumuamua's host star was ejected from its parental OC, we have no indication for the best case to consider: extrapolation of field stars AVR or consideration of OC AVR for younger ages. Therefore, following a conservative approach, we take the most extreme upper and lower limits obtained in both scenarios and report the age range of 'Oumuamua to be between 0.01--1.87 Gyr. Also, if an additional uncertainty of 20\% (from not knowing the conditions of 'Oumuamua ejection) is included, this constrained age range increases to 0--2.14 Gyr. The comparison between the ages obtained for both scenarios, and the combined result is shown in Figure \ref{fig::age_comparison}.

\section{Conclusions}
\label{sec::conclusions}

Based on dynamical arguments, we suggest that \designations\, will follow the same evolution of galactic orbits as that of stars. It is then expected that the same relations between kinematical parameters and age that we observe for star will also be observed for \designations. In this work, we have compared the orbital parameters derived for 'Oumuamua with the orbital parameters of the GCS as a function of age, to obtain kinematical constraints for the age of this \designation.

We performed orbital integrations using the module \texttt{galpy} considering 8 different models for the Galaxy (two potentials, two sets of $R_0$ and $v_0$ and two sets of $UVW_\odot$). The estimated orbital parameters range are $e = 0.043$--$0.076$, $z_\mathrm{max} = 27.05$--$27.91$ pc, $R_\mathrm{min} = 6.89$--$7.53$ kpc, $R_\mathrm{max} = 8.00$--$8.23$ kpc. We find that the orbit is then very close to planar and circular, which as noted by \citet{Gaidos+2017} is an indication of a lower age. However, in this work a more robust statistical approach was used to characterize the age of 'Oumuamua.

By performing a Bayesian analysis using the relation between velocity distribution and age, we estimated the age pdf for 'Oumuamua from its $UVW$ velocities. We have taken into account the bias in the point estimators, considered different values for the Solar motion, analysed the possible bias caused by the unknown ejection velocity from parental stellar system and investigated two different models of AVR. Finally, we estimated the age of $t_\mathrm{kin} = 0.50^{+1.37}_{-0.27}$ Gyr (with a most likely age of 0.45 Gyr and an expected age of 0.66 Gyr), when considering the extrapolated AVR for younger stars; and $t_\mathrm{kin} = 0.18^{+0.70}_{-0.17}$ (with a most likely age of 0.20 Gyr and an expected age of 0.11 Gyr) when considering the AVR of OCs for younger stars. From this estimations we constrained the possible age range for 'Oumuamua to be between 0.01--1.87 Gyr (0--2.14 Gyr, when considering the predicted additional uncertainty from 'Oumuamua's ejection process). The obtained age shows that, indeed, the kinematics of the object predict that it comes from a young stellar system, which may further constrain the list of parental system candidates.

During our analysis, we considered two models for the AVR, which differs for younger stars. In one of them, we extrapolate the AVR of field stars older than $\approx$ 1 Gyr, as done in A\&R18. For the other model, we considered that young stars are part of OCs and used the DAML sample to fit the AVR for these ages. We find that the slope of the AVR for OCs is smaller than for field stars and that the correlation between the $U$ and $V$ velocities can be neglected in this case. In both cases, the upper limit for the age of 'Oumuamua is well constrained, indicating that the object in fact had a space velocity of a younger body, before its encounter with the Sun.

As more objects like 'Oumuamua are observed (\citealp{PortegiesZwart+2017}, predict 2 to 12 encounters each year), we believe this method to derive their ages will further help the understanding of the characteristics of these objects and what they imply for proto-planetary disk formation.

\section*{Acknowledgements}

For providing support with a PhD grant, Almeida-Fernandes F. wants to thank CAPES --- The Brazilian Federal Agency for Support and Evaluation of Graduate Education within the Ministry of Education of Brazil. We would also like to thank the anonymous referee for the valuable comments that certainly helped to improve the quality of this work.




\bibliographystyle{mnras}
\bibliography{bibliography} 




\bsp	
\label{lastpage}
\end{document}